\documentclass[11pt]{article}
\usepackage{amsmath,amssymb,amsthm,amsxtra,overpic,bbm,bm,epsfig,ulem,color,multirow}
\usepackage{braket}

\begin{document}

\begin{center}
{\Large Flavon assisted low scale leptogenesis}
\end{center}

\vspace{0.05cm}

\begin{center}
{\bf Yan Shao, \bf Zhen-hua Zhao\footnote{Corresponding author: zhaozhenhua@lnnu.edu.cn}} \\
{ $^1$ School of Physics and Electronic Technology, Liaoning Normal University, Dalian 116029, China \\
$^2$ Center for Theoretical and Experimental High Energy Physics, \\ Liaoning Normal University, Dalian 116029, China }
\end{center}

\vspace{0.2cm}

\begin{abstract}
Low-scale leptogenesis scenarios, such as the resonant leptogenesis, typically require a highly degenerate mass spectrum of right-handed neutrinos (RHNs). This requirement can be circumvented by extending the seesaw framework with a scalar singlet $S$ that couples to RHNs via the $S N^{}_I N^{}_J$ terms (with $I \neq J$), which opens up new decay channels $N^{}_I \to N^{}_J S$ and provides additional sources of CP violation, thereby enabling successful leptogenesis at the TeV scale without the need for mass degeneracy. In this work, for the first time, we point out that the flavon fields, which are introduced in many flavor-symmetry neutrino mass models to be responsible for the generation of RHN masses through the acquisition of non-zero vacuum expectation values, serve as ideal candidates for the $S$ field. Taking as an example a flavor-symmetry neutrino mass model that naturally realizes the experimentally allowed TM1 mixing pattern and has the attractive features that only one flavon field plays the role of $S$ and that it couples to only two RHNs, we demonstrate that the observed neutrino masses and mixing angles can be consistently reproduced, while the observed baryon asymmetry can be achieved within a parameter space compatible with current experimental constraints.
\end{abstract}

\newpage

\section{Introduction}

The Standard Model (SM) of particle physics has achieved remarkable successes in describing most of the experimental observations. Nevertheless, it fails to account for two fundamental phenomena: the origin of neutrino masses and the observed baryon-antibaryon asymmetry of the Universe (BAU). On the one hand, neutrino oscillation experiments have established that neutrinos are massive and mixed states \cite{xing}, in clear contradiction with the SM prediction of massless neutrinos. On the other hand, cosmological observations indicate a tiny but non-vanishing excess of baryons over antibaryons, quantified by \cite{planck}
\begin{eqnarray}
Y^{}_{\rm B} \equiv \frac{n^{}_{\rm B}-n^{}_{\rm \bar B}}{s} \simeq (8.69 \pm 0.04) \times 10^{-11} \;,
\label{1.1}
\end{eqnarray}
where $n^{}_{\rm B}$ ($n^{}_{\rm \bar B}$) denotes the baryon (antibaryon) number density and $s$ is the entropy density.

A particularly appealing framework that can address both the neutrino mass and BAU puzzles is the type-I seesaw model \cite{seesaw1}-\cite{seesaw5}, which extends the SM by introducing heavy right-handed neutrinos (RHNs) $N^{}_I$ ($I=1, 2, 3$). The Yukawa interactions between left- and right-handed neutrinos generate a Dirac mass matrix $M^{}_{\rm D} = Y^{}_{\nu} v$, where $Y^{}_{\nu}$ is the neutrino Yukawa coupling matrix and $v = 174~{\rm GeV}$ is the Higgs vacuum expectation value (VEV). The RHNs also possess a Majorana mass matrix $M^{}_{\rm R}$. Under the seesaw condition $M^{}_{\rm R} \gg M^{}_{\rm D}$, the effective light-neutrino mass matrix is given by
\begin{eqnarray}
M^{}_{\nu} = - M^{}_{\rm D} M^{-1}_{\rm R} M^{T}_{\rm D} \;.
\label{1.2}
\end{eqnarray}
In this framework, the smallness of light neutrino masses follows naturally from the large Majorana masses of the RHNs. Moreover, the same setup provides an elegant mechanism for generating the BAU via leptogenesis \cite{leptogenesis}-\cite{Lreview4}.

However, the traditional type-I seesaw mechanism and  associated leptogenesis mechanism have the drawback that the newly introduced particle states (i.e., RHNs) are too heavy to be directly accessed by foreseeable experiments: on the one hand, in order to naturally generate sub-eV neutrino masses with $\mathcal{O}(1)$ Yukawa couplings, the RHN masses need to be around ${\cal O}(10^{13})$ GeV; on the other hand, when the RHN masses are hierarchical, in order to generate sufficient baryon asymmetry, the lightest RHN mass needs to be above ${\cal O}(10^9)$ GeV \cite{DI}. This motivates the search for alternative scenarios that can generate the observed baryon asymmetry at lower energy scales. Among the widely studied scenarios are resonant leptogenesis \cite{resonant1,resonant2}, where quasi-degenerate RHNs lead to a resonant enhancement of CP violation for the out-of-equilibrium decays (freeze-out) of RHNs, and leptogenesis via neutrino oscillations proposed by Akhmedov, Rubakov and Smirnov (ARS) \cite{ARS1,ARS2}, in which the baryon asymmetry originates from the out-of-equilibrium productions (freeze-in) of RHNs. These scenarios demonstrate that viable leptogenesis can be realized well below the conventional high scale. However, both scenarios typically require a highly degenerate RHN mass spectrum, which may need additional theoretical justification.

A simple and elegant alternative scenario is to introduce a scalar singlet $S$ which couples to the heavy Majorana neutrinos via the terms $SN^{}_I N^{}_J$ \cite{LeDall:2014too}-\cite{Ghosh:2024mpz}. These terms newly open up the decay channels $N^{}_I \to N^{}_J S$ (for $I \neq J$), which can enhance the departure from thermal equilibrium and introduce new sources of CP violation at the one-loop level. As a result, successful leptogenesis can be achieved even with TeV-scale RHNs, without the need for a highly degenerate mass spectrum. Such a scalar singlet can arise naturally in many well-motivated extensions of the SM, for instance as a dark matter candidate~\cite{Silveira:1985rk}-\cite{GAMBIT:2017gge}, an inflaton~\cite{Lerner:2009xg}-\cite{Ema:2017ckf}, or a pseudo-Nambu-Goldstone boson linked to the origin of the Higgs potential~\cite{Alanne:2017sip, Abe:2021mfy}.

We point out, for the first time, that the flavon fields ubiquitously present in flavor-symmetry models of neutrino masses naturally serve as the scalar singlet $S$ required for low-scale leptogenesis \footnote{In this paper, the flavon appears as a decay product of RHNs, whereas Ref.~\cite{Chen:2019wnk} studied the possibility that the flavon itself may directly generate the lepton asymmetry through its own decays. }. This establishes a previously unrecognized link between two active research directions (i.e., flavor-symmetry neutrino mass models and low-scale leptogenesis). In the specific realization studied in this work, the underlying flavor symmetry strongly restricts the flavon interactions with the RHNs. In the flavor basis, only one flavon interaction involving two RHNs is allowed. This interaction both modifies the RHN mass matrix to generate the TM1 neutrino mixing structure and provides the couplings relevant for leptogenesis. The model therefore serves as a minimal and highly constrained realization of low-scale leptogenesis subject to the restrictions imposed by the flavor symmetry. The successful realization of flavon-assisted low-scale leptogenesis under these restrictive conditions is non-trivial and also suggests that more general flavor-symmetry models with additional flavon interactions may provide a larger viable parameter space for this mechanism. This is because in many flavor-symmetry neutrino mass models~\cite{FS1}--\cite{FS4}, flavon fields are introduced to generate RHN masses through the acquisition of non-zero VEVs. To illustrate this point, we take the classic $A^{}_4$ flavor-symmetry neutrino mass model proposed in Ref.~\cite{A4} as an example. In this model, the three RHNs $N$ are arranged into a triplet representation of the $A^{}_4$ group and are assigned a charge of $\omega^2$ (where $\omega$ is the cubic root of unity) under an auxiliary $Z^{}_3$ group. Under this setup, the RHN masses are forbidden in the symmetry limit. Subsequently, two flavons, $\xi$ and $\varphi^{}_S$ -- which transform as the identity and triplet representations of $A^{}_4$, respectively, and both carry a charge $\omega$ under $Z^{}_3$ -- generate the desired RHN masses through the terms $\xi NN$ and $\varphi^{}_S NN$ after acquiring non-zero VEVs and spontaneously breaking the flavor symmetry. Furthermore, due to the special alignment of the VEVs of the three components of $\varphi^{}_S$, the resulting neutrino mixing follows the famous tri-bimaximal (TBM) pattern \cite{TB1, TB2-TM}. More importantly, for the purpose of the present paper, this mechanism naturally yields the terms $\xi N^{}_I N^{}_J$ and $\varphi^{}_S N^{}_I N^{}_J$, with $\xi$ and $\varphi^{}_S$ serving as $S$.

While the above $A^{}_4$ model illustrates the general idea, it predicts the now-disallowed TBM mixing (which predicts a vanishing neutrino mixing angle $\theta^{}_{13}$). Therefore, in this paper we instead take the flavor-symmetry neutrino mass model proposed in Ref.~\cite{Zhao:2014yaa} by one of the present authors as an example to illustrate that flavons can facilitate low-scale leptogenesis without requiring a highly degenerate RHN mass spectrum. This model naturally realizes the trimaximal (TM) mixing patterns~\cite{TB2-TM, TM-1}--\cite{TM-5} and is consistent with current experimental results. In this construction, the flavon field plays a dual role by connecting the realization of the TM1 mixing structure with the leptogenesis mechanism considered in this work. Furthermore, since the leptogenesis scale considered in this work lies in the regime where flavor effects can affect the evolution of the generated asymmetries, we incorporate the flavor-dependent Boltzmann evolutions of the lepton asymmetries in our analysis. This provides a more complete description of the generation of the baryon asymmetry.
As we will see in the next section, this model also has the attractive features that only one flavon field can play the role of $S$ and it only couples with two RHNs, which simplify the analysis.

The remaining parts of this paper are organized as follows. In the next section, we first briefly introduce the flavor-symmetry neutrino mass model of Ref.~\cite{Zhao:2014yaa} as a basis of our study. In Section~3, we discuss in detail the correspondingly modified Boltzmann equations, leading to the successful generation of the observed baryon asymmetry. In section~4, we systematically analyze the viable parameter regions of the model, taking into account constraints from other experimental measurements. Finally, the summary of our main results will be given in section~5.

\section{A brief introduction to the model}

In this section, we first give a brief introduction to the flavor-symmetry neutrino mass model of Ref.~\cite{Zhao:2014yaa}, including its motivation and key ingredients.

In the literature, one of the most popular approaches to realizing a particular neutrino mixing pattern is the so-called indirect approach~\cite{FS2}. In this approach, a given neutrino mixing pattern is realized with the help of certain flavor symmetries by ensuring that the RHN mass matrix $M^{}_{\rm R}$ be diagonal and that the three columns of the Dirac neutrino mass matrix $M^{}_{\rm D}$ be respectively proportional to the three columns of the neutrino mixing matrix $U$.
To be specific, the TBM mixing pattern
\begin{eqnarray}
U^{}_{\rm TBM}= \frac{1}{\sqrt 6} \begin{pmatrix}
-2 & \sqrt{2} & 0 \\
1 &  \sqrt{2}  & -\sqrt{3}  \\
1 &  \sqrt{2}  & \sqrt{3}
\end{pmatrix} \;,
\label{2.1}
\end{eqnarray}
can be naturally obtained when the three columns of $M^{}_{\rm D}$ are respectively proportional to the three columns of $U^{}_{\rm TBM}$ while $M^{}_{\rm R}$ is diagonal.

As mentioned above, the observation of a sizable $\theta^{}_{13}$ requires modifications to the TBM mixing pattern. Among the viable alternatives, the TM1 mixing pattern $U^{}_{\rm TM1}$ -- which preserves the first column of $U^{}_{\rm TBM}$ while modifying the other two -- provides a simple and predictive framework consistent with current experimental results~\cite{global1,global2}. Moreover, the TM1 mixing pattern has received further support from precision measurements by JUNO~\cite{JUNO:2025gmd} and has attracted renewed interest recently~\cite{Xing:2025bdm}--\cite{Ardakanian:2026rwz}.

From a phenomenological point of view, $U^{}_{\rm TM1}$ can be expressed as a product of $U^{}_{\rm TBM}$ and a unitary rotation $V$ on the 23-plane (i.e., $U^{}_{\rm TM1} = U^{}_{\rm TBM} V$). Motivated by this observation, Ref.~\cite{Zhao:2014yaa} constructs a flavor-symmetry neutrino mass model that realizes the TM1 mixing pattern via a minimal modification to the model that yields the TBM mixing pattern through the indirect approach described above.
The model is based on the flavor symmetries $A^{}_4 \times Z_2^1 \times Z_2^2 \times Z_2^3$. The $Z_2^I$ symmetries are auxiliary symmetries analogous to the $Z^{}_3$ symmetry in Ref.~\cite{A4}. Three right-handed neutrinos $N_I^0$ (here the superscript $0$ labels the RHN fields in the flavor basis, while $N^{}_I$ denotes the RHN fields in their mass basis, as will be specified below) are singlets under $A^{}_4$ and each is odd under its corresponding $Z_2^I$ symmetry. Under this setup, the right-handed neutrino mass matrix is initially constrained to be diagonal
\begin{equation}
M^{0}_{\rm R}= \operatorname{diag}(M^{0}_1, M^{0}_2, M^{0}_3) \;.
\label{2.2}
\end{equation}
Meanwhile, the three columns of the Dirac neutrino mass matrix are respectively proportional to the three columns of $U^{}_{\rm TBM}$:
\begin{equation}
M^{0}_{\rm D} = v\, U^{}_{\rm TBM}
\begin{pmatrix}
y^{}_1 & 0 & 0 \\
0 & y^{}_2 & 0 \\
0 & 0 & y^{}_3
\end{pmatrix} \;,
\label{2.3}
\end{equation}
where $y^{}_i$ are dimensionless parameters.
With this construction, the TBM mixing pattern is naturally realized.

To achieve the TM1 mixing, the key ingredient of this model is the introduction of a flavon field $S$ (denoted as $\xi$ in Ref.~\cite{Zhao:2014yaa}), which is odd under both $Z_2^2$ and $Z_2^3$ so that it can have a cross coupling between $N_2^0$ and $N_3^0$ of the form $\alpha^{}_0 S N_2^0 N_3^0$ with $\alpha^{}_0$ being a dimensionless coefficient.
After acquiring a non-zero VEV $v^{}_{S}$, this flavon field induces a mixed mass term $\Delta M = \alpha^{}_0 v^{}_{S}$ between $N_2^0$ and $N_3^0$, leading the right-handed neutrino mass matrix to take the form
\begin{equation}
M_{\rm R}^{\prime} = \begin{pmatrix}
M_1^0 & 0 & 0 \\
0 & M_2^0 & \Delta M \\
0 & \Delta M & M_3^0
\end{pmatrix} \;.
\label{2.4}
\end{equation}
While the Dirac neutrino mass matrix retains the form in Eq.~(\ref{2.3}), it is this $\Delta M$ term that induces the unitary rotation on the 23-plane which is needed to modify the TBM mixing pattern into the TM1 mixing pattern. Importantly, this $2$--$3$ rotation originates from the same interaction
allowed by the flavor symmetry that enters the leptogenesis dynamics. Thus, the RHN mixing responsible for the TM1 structure and the interactions relevant for leptogenesis are not independent ingredients of the model.
In Ref.~\cite{Zhao:2014yaa}, CP symmetry is further introduced, which constrains $\alpha^{}_0$ to be a real parameter and is spontaneously broken by the purely imaginary value of $v^{}_{S}$. In the present paper, for generality, we will not impose CP symmetry, so $\alpha^{}_0$ is treated as an arbitrary complex parameter.
Consequently, $v^{}_{S}$ is not treated as an independent parameter in our analysis, as its effect is encoded in the induced mass term $\Delta M$.

In order to obtain the RHN masses and subsequently facilitate the leptogenesis calculations, we go into the mass basis of RHNs. This is done by transforming the right-handed neutrino mass matrix in Eq.~(\ref{2.4}) into a diagonal form via a unitary rotation $V$ on the 23-plane:
\begin{eqnarray}
V^{T} M^{\prime}_{\rm R} V = {\rm diag}(M^{}_1, M^{}_2, M^{}_3) \;,
\label{2.5}
\end{eqnarray}
where $M^{}_I$ ($I=1, 2, 3$) denote the physical masses of the three RHNs. It is easy to see that $M^{}_1 = M^{0}_1$ and
\begin{eqnarray}
M^{}_{2} &=& \frac{1}{2} \left| M^{0}_{2} + M^{0}_{3} - \sqrt{(M^{0}_{2} - M^{0}_{3})^2 + 4(\Delta M)^2} \right| \;, \nonumber \\
M^{}_{3} &=& \frac{1}{2} \left| M^{0}_{2} + M^{0}_{3} + \sqrt{(M^{0}_{2} - M^{0}_{3})^2 + 4(\Delta M)^2} \right| \;.
\label{2.6}
\end{eqnarray}
This means that the mass eigenstate $N^{}_1$ coincides with $N^{0}_1$, whereas $N^{}_2$ and $N^{}_3$ are certain superpositions of $N^{0}_2$ and $N^{0}_3$. Meanwhile, the Dirac neutrino mass matrix in Eq.~(\ref{2.3}) is transformed into
\begin{eqnarray}
M^{}_{\rm D} = M^{0}_{\rm D} V \;.
\label{2.7}
\end{eqnarray}
Furthermore, after the basis transformation from $N^{0}_I$ to $N^{}_I$, the term $\alpha^{}_0 S N^{0}_2 N^{0}_3$ yields the terms $\alpha^{}_{IJ} S N^{}_I N^{}_J$ (for $I, J = 2, 3$) with
\begin{eqnarray}
\alpha^{}_{22} = 2\alpha^{}_0 V^{}_{32} V^{}_{22} \;, \qquad
\alpha^{}_{33} = 2\alpha^{}_0 V^{}_{33} V^{}_{23} \;, \qquad
\alpha^{}_{23} = \alpha^{}_{32} = \alpha^{}_0 \bigl( V^{}_{22} V^{}_{33} + V^{}_{23} V^{}_{32} \bigr) \;,
\label{2.8}
\end{eqnarray}
where $V^{}_{IJ}$ denotes the $IJ$-th component of $V$.
Among these terms, the off-diagonal coupling $\alpha^{}_{23} S N^{}_2 N^{}_3$ plays a crucial role for the purpose of this paper.

One can see that the model introduced above has three notable features. First, it achieves the TM1 mixing pattern by minimally modifying the model that realizes the TBM mixing pattern. Second, only one flavon field participates in the generation of RHN masses and their mixing, and thus plays the role of $S$ that opens up the decay channels $N^{}_I \to N^{}_J S$. Third, the flavor symmetry restricts the flavon interaction to $N_2$ and $N_3$, and the resulting couplings are all determined by the same rotation matrix $V$. This makes the present model a minimal and highly constrained realization of low-scale leptogenesis assisted by flavons, without requiring a highly degenerate RHN mass spectrum.

\section{Consequence for leptogenesis}

In this section, we study the consequences of the model introduced in the previous section for leptogenesis. In particular, we show that the flavon field $S$ can assist the realization of low-scale leptogenesis without requiring a highly degenerate RHN mass spectrum. Previous studies on scalar-assisted leptogenesis have mainly adopted the unflavored treatment for the evolution of the lepton asymmetry. In this work, we extend the analysis by including the flavor-dependent evolution of the lepton asymmetries, which provides a more complete description of the low-scale leptogenesis dynamics. In the following discussion, we assume that $N^{}_1$, which decouples from the other two RHNs, is the heaviest one so that its effects on leptogenesis can be safely neglected, and take $M^{}_3 > M^{}_2$ (in which case the decay channel $N^{}_3 \to N^{}_2 S$ is kinematically allowed for appropriate values of $m^{}_S$, the mass of $S$) without loss of generality.

In the conventional leptogenesis scenario, the amount of lepton-antilepton asymmetry produced crucially depends on the CP asymmetries between the RHN decay processes $N^{}_I \to L^{}_\alpha  H$ (for $\alpha = e, \mu, \tau$) and their CP-conjugate processes $N^{}_I \to \overline{L}^{}_\alpha  \overline{H}$. In the case that the RHN masses are hierarchical, the flavor-specific CP asymmetries for the decays of $N^{}_I$ are given by
\begin{eqnarray}
&& \varepsilon^{0}_{I \alpha} = \frac{1}{8\pi (Y^\dagger_{\nu}
Y^{}_{\nu})^{}_{II}} \sum^{}_{J \neq I} \left\{ {\rm Im}\left[(Y^*_{\nu})^{}_{\alpha I} (Y^{}_{\nu})^{}_{\alpha J}
(Y^\dagger_{\nu} Y^{}_{\nu})^{}_{IJ}\right] {\cal F} \left( \frac{M^2_J}{M^2_I} \right) \right. \nonumber \\
&& \hspace{1.cm}
+ \left. {\rm Im}\left[(Y^*_{\nu})^{}_{\alpha I} (Y^{}_{\nu})^{}_{\alpha J} (Y^\dagger_{\nu} Y^{}_{\nu})^*_{IJ}\right] {\cal G}  \left( \frac{M^2_J}{M^2_I} \right) \right\} \; ,
\label{3.1}
\end{eqnarray}
with ${\cal F}(x) = \sqrt{x} \{(2-x)/(1-x)+ (1+x) \ln [x/(1+x)] \}$ and ${\cal G}(x) = 1/(1-x)$.

In the model considered here, thanks to the new interaction $S N^{}_2 N^{}_3$, the flavor-dependent CP asymmetries for the decays of $N^{}_3$ receive additional vertex and self-energy contributions mediated by $S$ (see Figure~1)~\cite{LeDall:2014too}:
\begin{eqnarray}
\varepsilon^{}_{3\alpha} = \varepsilon^{0}_{3\alpha} + \varepsilon^{v}_{3\alpha} + \varepsilon^{s}_{3\alpha} \;.
\label{3.2}
\end{eqnarray}
The terms $\varepsilon^{v}_{I\alpha}$ and $\varepsilon^{s}_{I\alpha}$ can be generally expressed as
\begin{eqnarray}
&&\varepsilon^{v}_{I\alpha}=\sum_{J}\left(\frac{\text{Im}\left\{(Y^*_{\nu})^{}_{\alpha J} (Y^{}_{\nu})^{}_{\alpha I}\beta\alpha_{IJ}\right\}}{8\pi
(Y^\dagger_{\nu} Y^{}_{\nu})_{II}M_I}\mathcal{F}^v_{IJ,R}+\frac{\text{Im}\left\{(Y^*_{\nu})^{}_{\alpha J} (Y^{}_{\nu})^{}_{\alpha I}
\beta\alpha_ { IJ }
^*\right\}}{8\pi(Y^\dagger_{\nu} Y^{}_{\nu})_{II} M_I}\mathcal{F}^v_{IJ,L}\right),
\nonumber \\
&&\varepsilon^{s}_{I\alpha}=\sum_{J, K}\Bigg(\frac{\text{Im}\{(Y^*_{\nu})^{}_{\alpha K} (Y^{}_{\nu})^{}_{\alpha I}\alpha_{KJ}\alpha_{IJ}^*\}}{8\pi
(Y^\dagger_{\nu} Y^{}_{\nu})_{II}}\mathcal{F}^s_{IJK,RL}+\frac{\text{Im}\{(Y^*_{\nu})^{}_{\alpha K} (Y^{}_{\nu})^{}_{\alpha I}\alpha_{KJ}^*\alpha_{IJ}^*\}}{
8\pi(Y^\dagger_{\nu} Y^{}_{\nu})_{II}}\mathcal{F}^s_{IJK,LL}
\nonumber \\
 &&\hspace{2.0cm} +\frac{\text{Im}\{(Y^*_{\nu})^{}_{\alpha K} (Y^{}_{\nu})^{}_{\alpha I}\alpha_{KJ}\alpha_{IJ}\}}{
8\pi(Y^\dagger_{\nu} Y^{}_{\nu})_{II}}\mathcal{F}^s_{IJK,RR}+\frac{\text{Im}\{(Y^*_{\nu})^{}_{\alpha K} (Y^{}_{\nu})^{}_{\alpha I}\alpha_{KJ}^*\alpha_{IJ}\}}{
8\pi(Y^\dagger_{\nu} Y^{}_{\nu})_{II}}\mathcal{F}^s_{IJK,LR}\Bigg),
\label{3.3}
\end{eqnarray}
with
\begin{eqnarray}
&&\mathcal{F}^v_{IJ,L}=\left(-\sqrt{\delta^{}_{JI}}+r_{JI}\ln G\right),\quad\mathcal{F}^v_{IJ,R}=\left(-\sqrt{r_{JI}}\sqrt{\delta^{}_{JI}}+\sqrt{r^{}_{JI}}\ln G\right), \nonumber \\
&&\mathcal{F}^s_{IJK,LR}=\frac{\sqrt{\delta^{}_{JI}}}{2}\frac{\sqrt{\delta^{}_{JI}+4r^{}_{JI}}}{1-r^{}_{KI}}, \quad\mathcal{F}^s_{IJK,RR}=\sqrt { \delta^{}_{JI} }\frac{\sqrt{r^{}_{JI}}\sqrt{r_{KI}}}{1-r^{}_{KI}},  \nonumber \\
&&\mathcal{F}^s_{IJK,LL}=\sqrt{\delta^{}_{JI}}\frac{\sqrt{r^{}_{JI}}}{1-r^{}_{KI}}, \quad\mathcal{F}^s_{IJK,RL}=\frac{\sqrt{ \delta^{}_{JI} } } {2}\frac{\sqrt{r_{KI}}\sqrt{\delta^{}_{JI}+4r^{}_{JI}}}{1-r^{}_{KI}}, \nonumber \\
&& G=\frac{\sqrt{\delta^{}_{JI}+4r_{JI}\sigma_I+2\sigma_I}+\sqrt {\delta^{}_{JI}}}{\sqrt{\delta^{}_{JI}+4r_{JI}\sigma_I+2\sigma_I}-\sqrt{\delta^{}_{JI}}} \frac{\sqrt{\delta^{}_{JI}+4r_{ji}\sigma_I}-\sqrt{\delta^{}_{JI}}}{\sqrt{ \delta^{}_{JI}+4r_{JI}\sigma_I}+\sqrt{\delta^{}_{JI}}} , \nonumber \\
&& r^{}_{IJ}=\frac{M_I^2}{M_J^2},\quad\sigma_I=\frac{m_S^2}{M_I^2},\quad\delta^{}_{IJ} =(1-r^{}_{IJ}-\sigma_J)^2-4r^{}_{IJ}\sigma^{}_J \;.
\label{3.4}
\end{eqnarray}
In Eq.~(\ref{3.3}), $\beta$ is the dimensional coupling parameter of the tri-linear term $\beta S |H|^2$. This term is indispensable for the appearance of the left diagram in Figure~1. Its origin will be discussed around Eq.~(\ref{4.1}). It is worth noting that the expression for the vertex contribution already incorporates the contribution from 3-body final states discussed in Ref.~\cite{LeDall:2014too}.

\begin{figure*}
\centering
\includegraphics[width=5.5in]{Figure1.pdf}
\caption{ Flavon $S$ mediated one-loop diagrams
that enhance the amount of $CP$ violation in the decay $N^{}_3 \to L^{}_\alpha  H$.  }
\label{fig1}
\end{figure*}

To describe the evolution of the flavor-dependent lepton asymmetries generated by RHN decays, we extend the Boltzmann equations formulated in Ref.~\cite{LeDall:2014too} by taking into account the flavor dependence of the source and washout terms. In addition to the standard decay and inverse decay processes $N^{}_I \leftrightarrow L^{}_\alpha H$, the transition $N^{}_3 \to N^{}_2 S$ as well as the dominant $\Delta L = 2$ scattering processes $N^{}_I N^{}_J \to HH$ mediated by $S$ are also included. The relevant evolution equations take the form
\begin{eqnarray}
&& \frac{\mathrm{d} N^{}_{N_2}}{\mathrm{d} z} = - (D_2 + D_{32})\, \Delta^{}_{N_2} + D^{}_{32}\, \Delta^{}_{N_3} -\Delta^{}_{N_2N_3}S^{}_{N_2N_3\rightarrow HH}-\Delta_{N_2N_2}S_{N_2N_2\rightarrow HH} \;,
\nonumber \\
&&  \frac{\mathrm{d} N^{}_{N_3}}{\mathrm{d} z} = - (D_3 + D_{32})\, \Delta^{}_{N_3} + D^{}_{32}\, \Delta^{}_{N_2}
  -\Delta^{}_{N_2N_3}S^{}_{N_2N_3\rightarrow HH} -\Delta^{}_{N_3N_3}S^{}_{N_3N_3\rightarrow HH} \;,
\nonumber \\
&& \frac{\mathrm{d} N^{}_{\alpha}}{\mathrm{d} z} =  \sum_{I}^{} \varepsilon^{}_{I\alpha}\, D^{}_{I} \Delta^{}_{N_I} - W_{\alpha} N^{}_{\alpha} \;.
\label{3.5}
\end{eqnarray}
Here the temperature-dependent quantities are defined as
\begin{eqnarray}
&&  \Delta^{}_{N_I}(z) = \frac{N^{}_{N_I}(z)}{N_{N_I}^{\text{eq}}(z)} - 1\;, \hspace{1cm} \Delta_{N_IN_J} = \frac{N_{N_I}N_{N_J}}{N_{N_I}^{\mathrm{eq}}N_{N_J}^{\mathrm{eq}}}-1 \;, \hspace{1cm}
  N_{N_I}^\text{eq}(z) = \frac{z_I^2}{2}\,\mathcal{K}_2(z_I) \;,
\nonumber \\
&&   D_I (z)= K_I\,z \,\frac{\mathcal{K}_1(z_I)}{\mathcal{K}_2(z_I)} N_{N_I}^{\text{eq}}(z)\; , \hspace{1cm}
  D_{32}(z) = K_{32} \,z\,\frac{\mathcal{K}_1(z_1)}{\mathcal{K}_2(z_1)} N_{N_3}^{\text{eq}}(z)\;,
\nonumber \\
&&  W_{\alpha}(z) = \sum_I \frac{1}{4}\,K_{I\alpha}\,z_I^3\,\mathcal{K}_1(z_I) \; ,
\label{3.6}
\end{eqnarray}
where $\mathcal{K}^{}_{1,2}(z)$ are the modified Bessel functions, $z\equiv M^{}_{\rm min}/T$ (with $M^{}_{\rm min} = M^{}_2$ for the model considered here) is the dimensionless inverse temperature, and $z^{}_I \equiv M^{}_I / T$ corresponds to the $I$-th RHN.
The decay parameters $K^{}_I$ are defined as
\begin{eqnarray}
K^{}_{I} \equiv \sum_{\alpha}^{} K^{}_{I\alpha} = \sum_{\alpha}^{} \frac{\Gamma^{}_{I\alpha}}{H(T=M^{}_I)} \;,
\label{3.7}
\end{eqnarray}
where the decay width is given by $\Gamma^{}_{I\alpha} = |(Y^{}_{\rm \nu})_{\alpha I} |^{2} M^{}_I/(8\pi)$, and $H(T) =1.66 \sqrt{g_*}\, T^2/M^{}_{\rm Pl}$ is the Hubble rate, with $g^{}_*$ being the number of relativistic degrees of freedom and $M^{}_{\rm Pl} = 1.22\times 10^{19}$ GeV the Planck mass.
In analogy with $K^{}_I$, $K^{}_{32}$ is defined as
\begin{eqnarray}
K^{}_{32}  \equiv \frac{\Gamma(N_3\to N_2 S)}{H(T=M_3)} \;,
\label{3.8}
\end{eqnarray}
with
\begin{eqnarray}
\Gamma(N_3 \to N_2 S) = \frac{\left|\alpha^{}_{32}\right|^2 M_3}{16 \pi} \left[ \left(1+ \sqrt{r^{}_{23}}\right)^2 - \sigma_3 \right] \sqrt{\delta^{}_{23}} \;.
\label{3.9}
\end{eqnarray}
In addition to the usual inverse decays, further washout effects arise from the $S$-mediated $\Delta L=2$ scattering processes $N^{}_I N^{}_J \to HH$. The corresponding scattering functions entering in the Boltzmann equations are expressed as
\begin{eqnarray}
S^{}_{N^{}_I N^{}_J\rightarrow HH}\equiv\frac{M^{}_I}{64\pi^2H(T=M^{}_I)} \times\int_{w^{}_{\mathrm{min}}}^{\infty}\mathrm{d}w\sqrt{w}K^{}_1(\sqrt{w})\hat{\sigma}_{N^{}_I N^{}_J\rightarrow HH}\left(\frac{w M_I^2}{z_I^2}\right) \;,
\label{3.10}
\end{eqnarray}
with $w=s/T^2$ and $s^{}_{\rm min}=(M^{}_I+M^{}_J)^2$. Using the notation $\delta(s,M^{}_J,M^{}_J)=(s-M_I^2-M_J^2)^2-4M_I^2M_J^2$, the corresponding reduced cross section can be written as follows~\cite{Alanne:2017sip}
\begin{eqnarray}
\hat\sigma(N^{}_IN^{}_J\rightarrow HH)=\frac{1}{s}\,\delta(s,M^{}_I,M^{}_J)\,\sigma_{N^{}_IN^{}_J\rightarrow HH}=\frac{|\alpha^{}_{IJ}|^2\beta^2}{8\pi}\sqrt{\delta(s,M^{}_J,M^{}_J)}\frac{s-(M^{}_I+M^{}_J)^2}{s(s-m_{ S}^2)^2} \;.
\label{3.11}
\end{eqnarray}
By solving the above set of evolution equations, one obtains the final value of $Y^{}_{L}=\frac{\sum_{\alpha}^{} N^{}_{\alpha}}{27\times7.04}$. The baryon asymmetry is then given by $Y^{}_{\rm B} = c Y^{}_{\rm L}$, where $c = -28/79$ denotes the conversion efficiency from the lepton asymmetry to the baryon asymmetry via sphaleron processes.

\begin{figure*}
\centering
\includegraphics[width=4.5in]{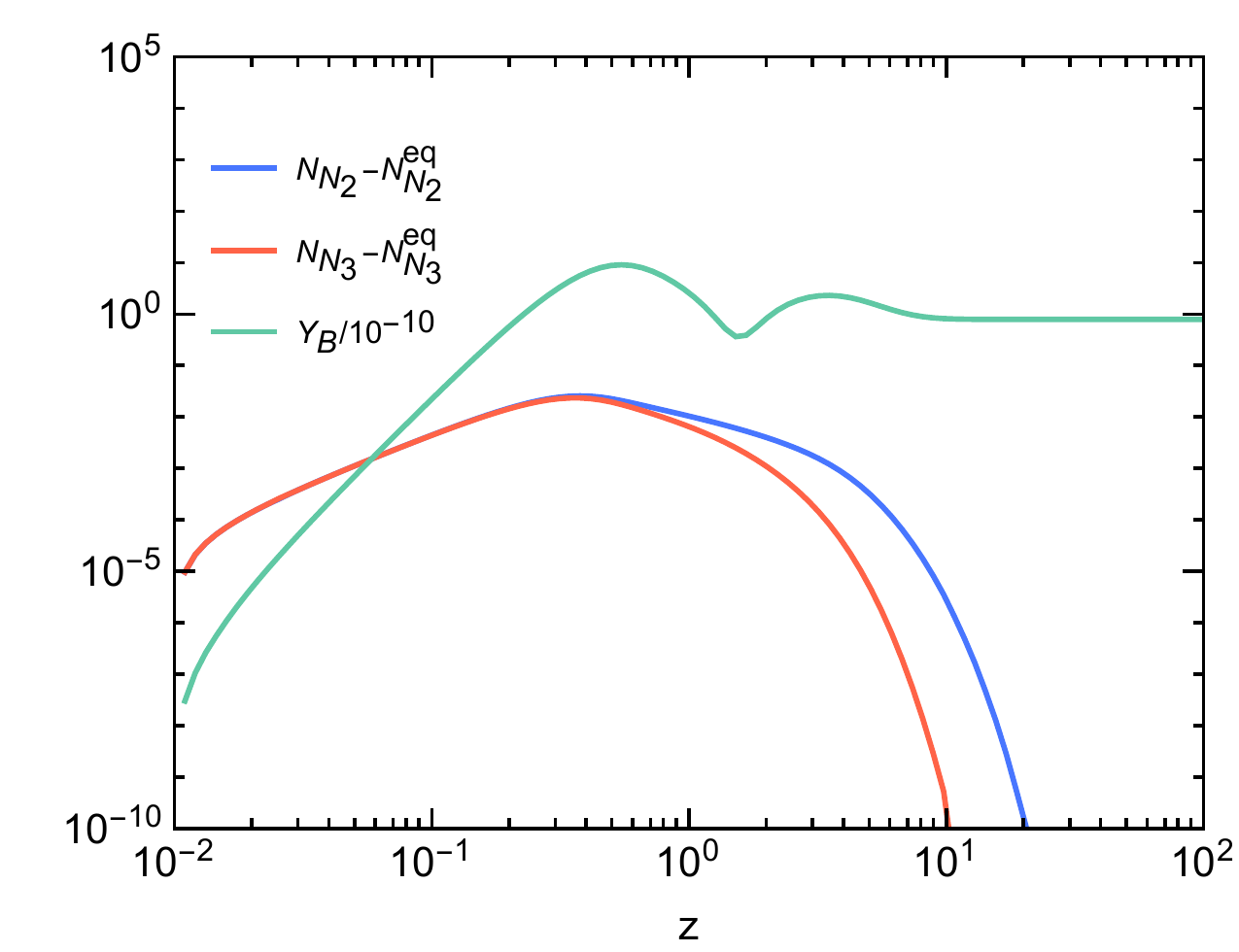}
\caption{ Evolutions of the RHN abundances $N^{}_{N^{}_2, N^{}_3}$ and the baryon asymmetry $Y^{}_{\rm B}$ as functions of $z$ for a representative parameter combination (see the main text for details). }
\label{fig2}
\end{figure*}

Now we are ready to perform the numerical calculations. Owing to the special forms of $M^{0}_{\rm D}$ and $M^\prime_{\rm R}$ in Eqs.~(\ref{2.3}) and~(\ref{2.4}), taking $M^{}_2$ and $M^{}_3$ as input parameters allows us to determine the coefficients $y^{}_i$ in Eq.~(\ref{2.3}) and the off-diagonal term $\Delta M$ in Eq.~(\ref{2.4}) from the experimental results for the neutrino parameters. These quantities provide the necessary inputs for solving the Boltzmann equations and computing the resulting baryon asymmetry. As an illustration, in Figure~\ref{fig2} we have shown the evolutions of the RHN abundances and the resulting baryon asymmetry as functions of $z$ for a representative parameter combination (with $M^{}_2 = 2$~TeV, $M^{}_3 = 4$~TeV, $m^{}_{S} = \beta = 1$~GeV, and $|\alpha^{}_{0}| = 10^{-1}$). At early times ($z \lesssim 1$), $N^{}_3$ departs from equilibrium, and its decays, including the $N^{}_3 \to N^{}_2 S$ channel, efficiently generate a lepton asymmetry. Around $z \sim 1$, the $N^{}_2$-related interactions become significant and partially wash out the previously generated asymmetry. The interplay between asymmetry generation from $N^{}_3$ decays and washout from $N^{}_2$ interactions yields a non-zero baryon asymmetry at late times.

\begin{figure*}
\centering
\includegraphics[width=6.5in]{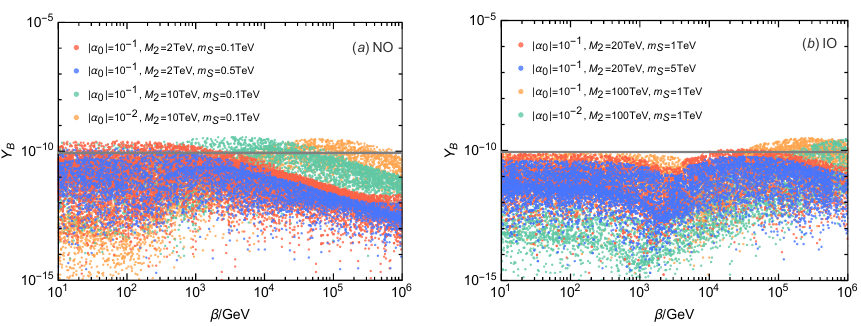}
\caption{ For the NO (a) and IO (b) cases of light neutrino masses, the allowed values of $Y^{}_{\rm B}$ as functions of $\beta$ for some benchmark values of $M^{}_2$, $m^{}_S$ and $|\alpha^{}_0|$. The horizontal line stands for the observed value of $Y^{}_{\rm B}$. }
\label{fig3}
\end{figure*}

In Figure~\ref{fig3}(a) and (b) (for the normal ordering (NO) and inverted ordering (IO) cases of light neutrino masses, respectively) we have shown the allowed values of $Y^{}_{\rm B}$ as functions of $\beta$ for some benchmark values of $M^{}_2$, $m^{}_S$ and $|\alpha^{}_0|$ (see the figure for details). These results are obtained for the following parameter settings: for the neutrino mass squared differences and neutrino mixing angles, we employ the global-fit results for them \cite{global1, global2}. For the two Majorana CP phases of the neutrino mixing matrix, we allow them to vary in the range 0---$2\pi$, while the Dirac CP phase is constrained by the TM1 mixing pattern realized by the model considered in the present paper. The lightest neutrino mass ($m^{}_1$ or $m^{}_3$ in the NO or IO case) is varied in the range between $0.001$ eV and $0.1$ eV. We take $|\alpha_0|=10^{-1}$ as a representative benchmark, since larger values of $|\alpha_0|$ generally allow a broader viable parameter region. We fix $M_3=1.3 M_2$ as a benchmark value. In order to realize the low-scale leptogenesis, the decay channel $N_3 \to N_2 S$ must be kinematically allowed, which requires $m^{}_{S} < M_3 - M_2$. When this channel is kinematically closed, the additional flavon-mediated vertex and self-energy contributions to the flavored CP asymmetries vanish, and the decay contribution $D_{32}$ is absent. For the representative parameter points presented in Figure~\ref{fig2}, the conventional flavored CP asymmetries in $N_3$ decays are only of order $10^{-16}-10^{-14}$, while the flavon-mediated contributions enhance the total asymmetries to the  $10^{-7}-10^{-5}$ level. Therefore, the $N_3\to N_2S$ channel is crucial for generating a sufficiently large CP asymmetry; without it, the conventional contribution is too small to account for the observed baryon asymmetry in the TeV-scale regime considered here.

The results show that in both the NO and IO cases the observed value of $Y^{}_{\rm B}$ can be successfully reproduced. In the NO case, for smaller values of $M^{}_2$ (e.g., $M^{}_2 = 2 $ TeV) and larger values of $|\alpha^{}_{0}|$ (e.g., $|\alpha^{}_{0}| = 10^{-1}_{} $), lower values of $\beta$ (up to around $10^3$ GeV) are needed to reproduce the observed value of $Y^{}_{\rm B}$ compared with the scenarios with larger $M^{}_2$. As $M^{}_2$ increases, the scale of $\beta$ needed to reproduce the observed $Y^{}_{\rm B}$ also rises.
In contrast, varying $m^{}_{S}$ has only a slight impact on the final baryon asymmetry. For the IO case, the required $\beta$ values are generally larger than those in the NO case.
Overall, the observed value of $Y^{}_{\rm B}$ can be successfully reproduced over a wide range of $\beta$ across the considered parameter space.

\section{Constraints from other experimental measurements}

In this section, we consider the constraints on the model parameters from other experimental measurements.

Since the $S$ field mixes with the Higgs doublet $H$ after electroweak symmetry breaking, its properties are subject to constraints from Higgs physics and collider searches. As $S$ is odd under the $Z^2_2$ and $Z^3_2$ symmetries of the model (see section~2 for details), the scalar potential relevant for $S$ and $H$ takes the following form:
\begin{eqnarray}
V(H,S) = m^2 |H|^2 +
\frac{1}{2}\mu^2_{}\,S^2 + \lambda^{}_{H} |H|^4+ \frac{1}{2}\lambda^{}_{SH}\,S^2|H|^2 +
\frac{1}{4}\lambda^{}_{S}\,S^4 \,,
\label{4.1}
\end{eqnarray}
which captures the leading interactions relevant for $H$-$S$ mixing and scalar mass generation. After electroweak and flavor symmetry breaking, the neutral scalar fields can be expanded around their vacuum expectation values as
\begin{eqnarray}
H=
\begin{pmatrix}
0\\
v+\dfrac{h}{\sqrt{2}}
\end{pmatrix} \;,
\qquad
S=v_S+s \;,
\label{4.2}
\end{eqnarray}
where $h$ and $s$ denote the fluctuations around the corresponding vacuum configurations. The minimization conditions of the scalar potential give
\begin{eqnarray}
m^2=-2\lambda_H v^2-\frac{1}{2}\lambda_{SH}v_S^2 \;,
\qquad
\mu^2=-\lambda_S v_S^2-\lambda_{SH}v^2 \;.
\label{4.3}
\end{eqnarray}
Accordingly, the scalar mass-squared matrix in the $(h,s)$ basis is
\begin{eqnarray}
\mathcal{M}_{\rm scalar}^2=
\begin{pmatrix}
4\lambda_H v^2 & \sqrt{2}\lambda_{SH}vv_S\\
\sqrt{2}\lambda_{SH}vv_S & 2\lambda_Sv_S^2
\end{pmatrix} \;.
\label{4.4}
\end{eqnarray}
The above mass matrix is diagonalized by an orthogonal rotation with a mixing angle $\theta$. We identify one of the mass eigenstates with the observed Higgs boson of mass $m_h$, and denote the mass of the second,
predominantly flavon-like scalar eigenstate by $m_S$. The quartic couplings can then be expressed in terms of the physical scalar parameters as
\begin{eqnarray}
&&  \lambda_H =
\frac{m_h^2\cos^2\theta+m_S^2\sin^2\theta}{4v^2} \;,
\nonumber \\
&&  \lambda_S =
\frac{m_h^2\sin^2\theta+m_S^2\cos^2\theta}{2v_S^2} \;,
\nonumber \\
&& \lambda_{SH}=
\frac{(m_S^2-m_h^2)\sin 2\theta}
{2\sqrt{2}\,v v_S} \;.
\label{4.5}
\end{eqnarray}
As discussed below Eq.~(\ref{3.4}), the trilinear coupling $\beta$ originates
from the portal interaction after $S$ acquires its VEV and satisfies
$\beta=\lambda_{SH}v_S$. Using the above relations, one obtains
\begin{equation}
\beta=
\frac{(m_S^2-m_h^2)\sin 2\theta}
{2\sqrt{2}\,v} \;.
\label{4.6}
\end{equation}
This relation makes explicit that $\beta$, $m_S$, and $\theta$ are correlated with each other rather than independent parameters. Moreover, the flavon VEV $v_S$ is further related to the RHN sector through $\Delta M=\alpha_0 v_S$, as discussed in section 2. Theoretical consistency imposes several constraints on the scalar potential. Vacuum stability requires $\lambda^{}_{H}, \lambda^{}_{S} > 0$ and $4\lambda^{}_{H}\lambda^{}_{S} - \lambda_{SH}^2 > 0$, ensuring that the potential is bounded from below. In addition, perturbativity demands that all quartic couplings remain within the perturbative regime. While a naive upper bound is $|\lambda^{}_i| < 4\pi$, we adopt the more conservative requirement $|\lambda^{}_i| < 3$, consistent with perturbative unitarity bounds~\cite{Goodsell:2018tti}.

In addition to these theoretical conditions, experimental constraints from Higgs measurements further restrict the parameter space. The mixing between the observed 125~GeV Higgs boson and the singlet scalar is constrained by Higgs signal-strength data, which imposes an upper bound on the mixing angle, $|\sin\theta| \lesssim 0.3$~\cite{Robens:2015gla, Ilnicka:2018def}. Moreover, if the singlet scalar is sufficiently light, i.e., $m^{}_{ S} < m^{}_h/2$, the decay channel $h \to SS$ becomes kinematically accessible and contributes to the invisible width of the Higgs boson. The corresponding partial decay width is given by
\begin{equation}
\Gamma_{h\rightarrow SS}=\frac{\lambda_{SH}^2 v_{ }^2}{16\pi m^{}_h} \sqrt{1-\frac{4m^2_S}{m^2_h}} \;,
\label{4.7}
\end{equation}
where $m^{}_h = 125$ GeV and the total SM Higgs width is $\Gamma_h^{\rm SM} \simeq 4.07$~MeV~\cite{LHCHiggsCrossSectionWorkingGroup:2011wcg}. The current experimental bound on invisible Higgs decays is $\mathrm{BR}(h\to \mathrm{inv}) \le 0.23$~\cite{ATLAS:2015ciy, CMS:2016dhk}.

\begin{figure*}
\centering
\includegraphics[width=6.5in]{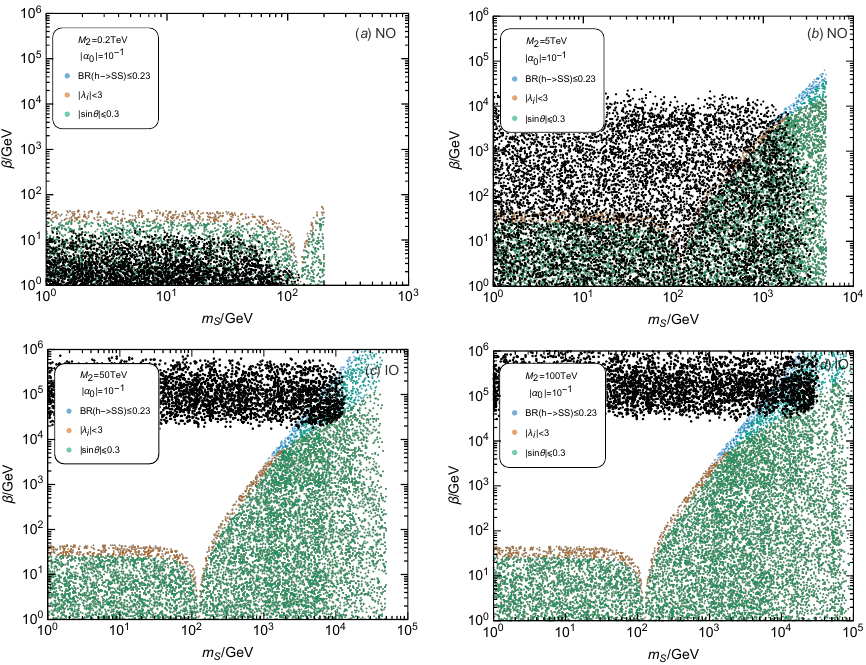}
\caption{
Viable parameter space of $\beta$ versus $m^{}_{S}$ for successful leptogenesis in the NO and IO cases of light neutrino masses. The black points denote the parameter region that reproduces the observed baryon asymmetry, while the colored points correspond to constraints from various experimental benchmarks.
}
\label{fig4}
\end{figure*}

Taking account of these constraints, in Figure~\ref{fig4} we have further shown the parameter space of $m^{}_{S}$ versus $\beta$ for successful leptogenesis, as well as the region consistent with current experimental constraints. The viable parameter space is identified as the overlap between these two regions. With the same parameter settings as in section~3, these results are obtained within the parameter space that allows for a successful reproduction of the observed baryon asymmetry of the Universe. To obtain a representative and sufficiently comprehensive parameter space, we vary the mass ratio $M_3/M_2$ within the range $1.3$--$2$, while taking $M_2$ and $|\alpha^{}_{0}|$ as the benchmark parameters.
The absence of points around $m_{S} \sim m_h$ originates from the near-degeneracy of the two scalar mass eigenstates. In this region, it becomes difficult to simultaneously satisfy the reality and perturbativity requirements imposed on the scalar potential parameters.
In the NO case, we find that $M^{}_{2} \sim 200 $ GeV corresponds to the lowest mass scale at which the observed $Y^{}_{\rm B}$ can be successfully reproduced within this parameter setup. At this scale, the viable parameter space is found for $\beta$ up to about $10$ GeV. As $M_2$ increases, larger values of $\beta$ become compatible with successful leptogenesis. For larger values of $M^{}_2$ (e.g., $M^{}_{2} = 5 $ TeV), successful leptogenesis is predominantly realized for $\beta$ values as large as $\sim 10$ TeV. In the IO case, a similar qualitative behavior is observed, but with a more restricted viable parameter space. To further illustrate the self-consistency of the scalar sector, we also examine the corresponding values of the remaining scalar parameters for the four benchmark scenarios shown in Figure~\ref{fig4}. The typical ranges of $v_S$, $\lambda_H$, $\lambda_S$, $\lambda_{SH}$, and $|\sin\theta|$ are summarized in Table~1, while all viable points satisfy the theoretical and experimental constraints discussed above.

\begin{table}
\centering
\begin{tabular}{cccccc}
\hline
Benchmark & $v_S$ & $\lambda_H$ & $\lambda_S$ & $\lambda_{SH}$ & $|\sin\theta|$ \\
\hline
Figure~\ref{fig4}(a) & $0.15$--$1.2$~TeV & $0.000026$--$0.13$ & $0.0029$--$0.18$ & $0.00046$--$0.017$ & $0.011$--$0.22$ \\
Figure~\ref{fig4}(b) & $3.3$--$45$~TeV & $0.000027$--$1.8$ & $0.0000019$--$0.052$ & $0.000013$--$0.46$ & $0.000019$--$0.3$ \\
Figure~\ref{fig4}(c) & $330$--$410$~TeV & $2.0$--$3.0$ & $0.000065$--$0.00029$ & $0.026$--$0.054$ & $0.043$--$0.089$ \\
Figure~\ref{fig4}(d) & $610$--$900$~TeV & $1.1$--$3.0$ & $0.000065$--$0.00045$ &$0.23$--$0.71$ & $0.013$--$0.047$ \\
\hline
\end{tabular}
\caption{Typical ranges of the scalar-sector parameters for the four benchmark scenarios shown in Figure~\ref{fig4}.}
\end{table}

Although the current Higgs-related searches constrain only part of the parameter space compatible with successful leptogenesis, several complementary possibilities remain for testing the scenario. The realization considered here corresponds to a particularly simple case in which the relevant flavon interaction, $\alpha_0 S N_2^0N_3^0$, induces
mixing only in the $N_2$--$N_3$ sector. Future precision measurements of the Higgs couplings, searches for exotic Higgs decays, and direct searches for additional scalar resonances can further probe the Higgs--flavon sector~\cite{Berger:2014gga}, while precision neutrino-oscillation measurements provide an independent test of the underlying TM1 flavor structure. More general flavor-symmetry constructions may involve several flavon fields and a more general RHN mixing structure~\cite{Luhn:2013lkn}. If these flavons communicate with the Higgs sector, the enlarged scalar sector may lead to additional scalar resonances and exotic or cascade scalar decays that can be tested at colliders~\cite{Robens:2019kga}. Moreover, heavier flavon-like states may decay into RHN pairs when kinematically allowed, providing an additional probe of the flavon--RHN sector~\cite{Fernandez-Martinez:2022stj}. A quantitative analysis of such more general realizations is model dependent and is left for future work.

\section{Summary}

The type-I seesaw model provides an elegant explanation for both neutrino masses and the baryon asymmetry of the Universe. However, in its traditional form, the newly introduced RHNs are too heavy to be directly accessible in foreseeable experiments. This has motivated low-scale leptogenesis scenarios such as resonant leptogenesis, but these typically require a highly degenerate RHN mass spectrum.
An alternative approach is to introduce a scalar singlet $S$ that couples to RHNs via the $S N^{}_I N^{}_J$ terms (with $I \neq J$), opening up new decay channels $N^{}_I \to N^{}_J S$ and providing additional sources of CP violation. This allows for TeV-scale leptogenesis without the need for a highly degenerate RHN mass spectrum.

In this paper, we have pointed out, for the first time, that the flavon fields ubiquitously present in flavor-symmetry models of neutrino masses naturally serve as the scalar singlet $S$ required for low-scale leptogenesis. This establishes a previously unrecognized link between two active research directions (i.e., flavor-symmetry neutrino mass models and low-scale leptogenesis).
We have taken the classic $A^{}_4$ model proposed in Ref.~\cite{A4} as an example to illustrate the general idea. Since the TBM mixing pattern predicted by this model has been ruled out by measurements of $\theta^{}_{13}$, we have instead adopted the model proposed in Ref.~\cite{Zhao:2014yaa} as an example for detailed numerical calculations. This model naturally realizes the experimentally allowed TM1 mixing pattern and has the attractive features that only one flavon field plays the role of $S$ and that it couples only to two RHNs (namely $N^{}_2$ and $N^{}_3$), which simplifies the analysis. This flavon field opens the new decay channel $N^{}_3 \to N^{}_2 S$, providing an additional source of CP violation and enhancing the departure from thermal equilibrium. In addition, we have incorporated flavor effects into the leptogenesis analysis by considering the flavor-dependent evolution of the generated lepton asymmetries, providing a more complete description of the resulting baryon asymmetry. Our analysis shows that the model consistently reproduces the observed neutrino masses and mixing while enabling viable leptogenesis at TeV scales without requiring degenerate RHNs.

Specifically, the analysis shows that the observed baryon asymmetry can be successfully reproduced in a significant region of parameter space consistent with current experimental constraints. In the NO case, successful leptogenesis is possible for $M^{}_{2}$ as low as $\sim 200$~GeV. At this scale, the viable parameter space is found for $\beta$ up to about $10$ GeV. As $M_2$ increases, larger values of $\beta$ become compatible with successful leptogenesis. For larger values of $M^{}_2$ (e.g., $M^{}_{2} = 5 $ TeV), successful leptogenesis is predominantly realized for $\beta$ values as large as $\sim 10$ TeV. In the IO case, a similar qualitative behavior is observed, but with a more restricted viable parameter space.

Before closing, we emphasize that although our analysis was carried out within a specific flavor-symmetry neutrino mass model, the underlying mechanism is nevertheless applicable to a broader class of similar models.

\vspace{0.5cm}

\underline{Acknowledgments} \vspace{0.2cm}

This work was supported in part by the National Natural Science Foundation of China under Grant Nos.~12475112 and 12447101, Liaoning Revitalization Talents Program under Grant No.~XLYC2403152, and the Basic Research Business Fees for Universities in Liaoning Province under Grant No.~LJ212410165050.

\end{document}